\documentstyle[12pt,epsfig,supercite]{article}

\newcommand{\GS}{\gamma_{\rm str}}
\newcommand{\BE}{\begin{equation}}
\newcommand{\EE}{\end{equation}}

\newcommand{\la}{\lambda}

\def\abstracts#1#2#3{{
        \centering{\begin{minipage}{4.62in}\baselineskip=13pt
        \small
          \centerline{\bf Abstract}
        \vspace*{0.2cm}                
        \parindent=0pt #1\par
        \parindent=18pt #2\par
        \parindent=15pt #3
        \end{minipage} }\par}}
\begin{document}
\vspace*{-4.5cm}
\hfill \parbox{5cm}{ ~\\~}\\               
\vspace*{1.0cm}
\centerline{\LARGE \bf Simplicial Quantum Gravity} \\[0.2cm]
\centerline{\LARGE \bf  on a  } \\[0.2cm]
\centerline{\LARGE \bf Randomly Triangulated Sphere} \\[0.5cm]
\centerline{\large {\em Christian Holm\/}$^{1}$ and
                   {\em Wolfhard Janke\/}$^{2,3}$}\\[0.4cm]
\centerline{\large    $^1$ {\small Max Planck Institut f\"ur Polymerforschung
                      }}
\centerline{    {\small Ackermannweg 10, 55128 Mainz, Germany }}\\[0.2cm]
\centerline{\large    $^2$ {\small Institut f\"ur Physik,
                      Johannes Gutenberg-Universit\"at Mainz}}
\centerline{    {\small Staudinger Weg 7, 55099 Mainz, Germany }}\\[0.2cm]
\centerline{\large    $^3$ {\small Institut f\"ur Theoretische Physik,
                      Universit\"at Leipzig}}
\centerline{    {\small Augustusplatz 10/11, 04109 Leipzig, Germany }}\\[0.5cm]
\abstracts{}
{
We study 2D quantum gravity on spherical topologies employing the Regge
calculus approach with the $dl/l$ measure. Instead of the normally used fixed
non-regular triangulation we study random triangulations which are generated
by the standard Voronoi-Delaunay procedure. For each system size we average
the results over four different realizations of the random lattices. We
compare both types of triangulations quantitatively and investigate how the
difference in the expectation value of the squared curvature, $R^2$, for fixed
and random triangulations depends on the lattice size and the surface area
$A$. We try to measure the string susceptibility exponents through finite-size
scaling analyses of the expectation value of an added $R^2$-interaction term,
using two conceptually quite different procedures. The approach, where an
ultraviolet cut-off is held fixed in the scaling limit, is found to be plagued
with inconsistencies, as has already previously been pointed out by us. In a
conceptually different approach, where the area $A$ is held fixed, these
problems are not present. We find the string susceptibility exponent
$\gamma_{\rm str}'$ in rough agreement with theoretical predictions for the
sphere, whereas the estimate for $\gamma_{\rm str}$ appears to be too
negative. However, our results are hampered by the presence of severe
finite-size corrections to scaling, which lead to systematic uncertainties
well above our statistical errors. We feel that the present methods of
estimating the string susceptibilities by finite-size scaling studies are not
accurate enough to serve as testing grounds to decide about a success or
failure of quantum Regge calculus. 
}
{}
\vspace*{0.5cm}
\thispagestyle{empty} 
\pagenumbering{arabic}
                  \section{INTRODUCTION}
Over the past few years Regge calculus has been extensively used in the study of 
quantum gravity \cite{q-regge}. In its usual form one
investigates regular simplicial triangulations of manifolds of a given topology, 
which up to now have been mostly hypertori. In this work we will study 
a model of two-dimensional quantum gravity where analytic 
calculations \cite{gamma_string} have shown that the internal fractal 
structure of the manifold depends very sensitively on the global topology. 
Important universal quantities are the string susceptibility exponent
$\gamma _{\rm str}$, which can be expressed in terms of the genus $g$ of the 
surface by the KPZ formula \cite{gamma_string}
$\gamma _{\rm str} = 2 - \frac{5}{2} (1-g)$, and the related exponent
$\gamma'_{\rm str}$, which is predicted \cite{kawai_r2} to take the value
$\gamma'_{\rm str} = 2 - 2(1-g)$. 

For the torus ($g = 1$) the Regge approach with the $dl/l$ measure 
has given results compatible with the KPZ formula, but this may well be a 
coincidence (note, e.g., that $\GS = \GS' = 2$ for $g=1$). For the sphere ($g = 0$) 
and topologies of higher gender, on the other hand, the situation is still 
unclear \cite{hamber91,bock,bock_proc,hj95a,hj95b}. One potential problem is 
that for the sphere only very small regular triangulations exist, such as
the tetrahedron, the octahedron, and the icosahedron. In order to obtain 
triangulations with a larger number of vertices, one either has to use 
{\em non-regular\/} triangulations or to resort to {\em random\/} 
triangulations.
Due to Euler's theorem
non-regular triangulations possess a certain number
of special vertices, which could delicately alter the finite-size scaling (FSS)
behavior. From this point of view it is, therefore, more appealing to use 
random triangulations where none of the vertices plays a special role and
which thus possess on the average the same properties.

In this work we use Monte Carlo (MC) simulations to study random 
triangulations of a sphere along the lines briefly reported in a recent
short communication \cite{hj97}. Here we present a detailed comparison
of our new results with our earlier estimates 
obtained by using the triangulated surface of a cube as spherical 
lattice \cite{hj95b}. We compute estimates of $\GS$ and the related 
exponent $\GS'$ on the random triangulations, employing both 
the scaling approach of Refs.~\cite{bock,bock_proc} and the novel FSS method 
proposed in Ref.~\cite{hj95a}. 

The remainder of the paper is organized as follows.
In Sec.~2 we define the model and the string susceptibility exponents. The
FSS predictions are briefly recalled in Sec.~3, and in Sec.~4 we give an
overview of the simulation set-up. The results are discussed in Sec.~5,
and in Sec.~6 we conclude with a summary and a few final remarks. 
%
            \section{MODEL}
%
When focusing on measurements of the string susceptibilities in Regge 
calculus, it is necessary to introduce a curvature square term in the action.
Under certain assumptions which will be presented in the next section, one
can derive a FSS expression for the expectation value of the curvature squared
term, which in turn allows one
to deduce an estimate of $\GS$ and $\GS'$ 
through FSS analyses. We therefore started out with the partition function
\begin{equation}
Z(A)=\!\!\int{\!\!{\cal D} \mu (q) e^{-\!\sum _i
{(\lambda A_i + a R_i^2)}} \delta (\sum _i A_i-A)},
\label{eq:1}
\end{equation}
where $R _i ^2 = \delta _i ^2/A_i$ denotes the local squared curvatures.
The $A_i$ are barycentric areas connected to the site $i$, defined as
\begin{equation}
A_i = \sum _{t \supset i} {\frac{1}{3}A_t},
\end{equation}
where $A_t$ denotes the area of the triangle $t$, and the
$\delta _i = 2 \pi - \sum_{t \supset i} {\theta _i (t)}$ are the
deficit angles, with $\theta_i (t)$ being the dihedral angle at vertex
$i$. The action contains the coupling constant $\lambda$ (the
cosmological constant), which is irrelevant in this case because of the
constant area constraint, and the coupling constant $a$ of the curvature
squared term, whose strength we
are going to vary.
The dynamical degrees of freedom of Regge calculus 
are the squared link lengths, 
$q=l^2$, which stand in a linear relation to the components of the metric 
tensor $g_{\mu\nu}$. 
Denoting by $g_{\mu\nu}(i)$ the components of the metric tensor for the 
$i^{th}$ triangle, and by $q_{i+\mu,i+\nu}, q_{i,i+\mu}$, and $q_{i,i+\nu}$
the square of its 
three edge lengths, one can derive that
$g_{\mu\nu}(i) = \frac{1}{2}\left[ q_{i,i+\mu} + 
q_{i,i+\nu} - q_{i+\mu,i+\nu}\right]$. 

The final important degree of freedom for quantum Regge calculus is the choice
of the functional integration measure.
We used the simple scale invariant 
``$dl/l$ computer measure''
${\cal D}\mu (q) = \left[ \prod _{\langle ij \rangle} \frac{dq_{ij}}{q_{ij}} 
\right] F_\epsilon(\{q_{ij}\})$, in order to produce results that are directly
comparable to those of our previous works \cite{hj95b,hj94a}, from which we also 
adopted the present notation. The function $F_\epsilon(\{q_{ij}\})$ ensures
that updates of the link lengths do not violate the triangle inequalities.
The proper choice of the functional measure 
is a very controversial issue which is actively analyzed in the current 
literature from an analytical point of view \cite{meno}. Since in the present
work we mainly focussed on a detailed investigation of different discretization 
schemes, a careful numerical analysis of the measure problem has to be 
postponed to future work.

In the partition function (\ref{eq:1}) the total area is constrained to 
be a constant such that the only dynamical term is the $R^2$-interaction. 
Its coupling constant $a$ sets an intrinsic length scale of $\sqrt{a}$, and 
$\hat A \equiv A/a$ can be used to distinguish between the cases of 
weak ($\hat A \gg 1$) and strong ($\hat A \ll 1$) $R^2$-gravity.
In the first limit the KPZ \cite{gamma_string} scaling behavior of the 
partition function is recovered, 
\begin{equation}
Z(A) \propto A^{\gamma_{\rm str} -3} e^{- \lambda _R A} \qquad 
\mbox{($\hat A \gg 1$)},
\end{equation}
where $\gamma_{\rm str} = 2 - \frac{5}{2} (1-g)$ is the string 
susceptibility exponent, with $g$ being the genus of the surface
under consideration, and $\lambda _R $ denotes the renormalized cosmological 
constant. Notice that the exponent $\gamma_{\rm str}$ appears only as
the sub-dominant correction to the large area behavior. This is one of the
reasons why numerical determinations of $\gamma_{\rm str}$ are so difficult.
In the opposite limit of strong $R^2$-gravity it
was found \cite{kawai_r2} that
\BE
Z(A) \propto A^{\gamma_{\rm str}' -3} e^{-S_c/\hat A} e^{- \lambda _R A -
b \hat A } \qquad \mbox{($\hat A \ll 1$)},
\EE
where $S_c = 16 \pi ^2(1-g)^2$ is the classical action, $b$ is some constant, 
and $\gamma_{\rm str}' = 2 - 2(1-g)$ is supposed to be another universal
exponent.

For later use we express the exponents in terms of the derivative of $Z$ 
with respect to $\hat A$:
\begin{eqnarray}
\frac{\partial \ln Z }{\partial \hat A}  &=&  
- a \lambda _R +
\frac{ \gamma _{\rm str} - 3}{\hat A} 
 \,~~~~~~~~~~~~~~~~~(\hat A \gg 1),
\label{ps.eq:lnZ} \\
\frac{\partial \ln Z }{\partial \hat A}  &=&  
S_c/\hat A^2
+ \frac{\GS ' - 3}{\hat A} - a \lambda_R - b 
 ~~~(\hat A \ll 1).
\label{ps.eq:lnZ_b}
\end{eqnarray}
The point is that the partition function $Z$ is not directly accessible in 
Monte Carlo simulations. Logarithmic derivatives as in
(\ref{ps.eq:lnZ}) and (\ref{ps.eq:lnZ_b}), on the other hand, are 
straightforward to estimate by measuring appropriate expectation values.

To discretize the global topology of a sphere we used random triangulations
that are constructed according to the Voronoi-Delaunay procedure, as described 
in Ref.~\cite{ranlat}. In this way we can control the influence of the special
vertices of non-regular triangulations. A sample random triangulation with 
$N_0 = 1500$ vertices is shown in Fig.~\ref{random_sphere.fig}. For spherical 
topologies we have the Euler relations $N_0 - 2 = N_1/3$, $N_0 - 2 =  N_2/2$, 
and $2N_1=3N_2$, where $N_0, N_1$, and $N_2$ denote the number of sites, links 
and triangles, respectively. In random triangulations the number of 
nearest neighbors $q$ varies, theoretically, between 3 and $\infty$. For
a finite number of vertices the total number of links is, of course, bounded 
from above, and in the practical realizations we typically find a maximal
coordination number of about $q_{\rm max} \approx 10-13$, depending on the
lattice size. Due to the global constraint on $N_0$ and $N_1$, the average
number of nearest neighbors in each realization is given by 
$\bar{q} = 6(1 - 12/N_0)$. For our largest realizations with $N_0 = 17\,498$
the probability distribution $P(q)$ of the coordination numbers $q$ is 
plotted in Fig.~\ref{fig:q_dist}. 
%
                   \section{FINITE-SIZE SCALING}
%
The previously used methods of Ref.\cite{hamber91,bock,bock_proc} to extract
$\GS$  
are plagued by inconsistencies, as has been discussed in our earlier 
work \cite{hj95b}. There we also suggested an alternative method to determine
the string susceptibilities. For the easier digestion of the present article
we recapitulate here only the important points. 
From the partition function (\ref{eq:1}) one can easily compute the derivative
\BE
\frac{\partial \ln Z}{\partial \hat A} = -a \lambda +
\frac{1}{\hat A}( \hat R^2 - 1),
\label{ps.eq:log}
\EE
where 
\BE
\hat R^2 := a\langle \sum _i R_i ^2 \rangle = 
 a\langle \sum _i \delta _i ^2 / A_i \rangle
\EE
is the scaled total curvature squared. By inspecting 
(\ref{eq:1}) one can then show that $\hat R^2 = \hat R^2(\hat A, N_1)$ 
depends only on $N_1$ and the dimensionless parameter $\hat A$. 

By comparing with (\ref{ps.eq:lnZ}) and (\ref{ps.eq:lnZ_b}) one expects
that for large $N_1$ the finite part of $\hat R^2$ can be expanded into 
power series, 
\begin{eqnarray}
\hat R^2 &=& b_1 \hat A + b_0 + b_{-1} / \hat A + \dots \qquad (\hat A \gg 1),
\label{eqn:R-infty} \\
\hat R^2 &=& b'_{-1} / \hat A + b'_0 + b'_1 \hat A + \dots \qquad 
(\hat A \ll 1),
\label{eqn:R-inftyb} 
\end{eqnarray}
with
\BE
b_1 = -a (\la _R - \la ), \quad b_0 = \GS - 2 \qquad \qquad \qquad \qquad
~~~(\hat A \gg 1), 
\EE
and
\BE
b'_{-1} = S_c, \quad
b'_0 = \GS ' - 2,  \quad
b'_1 = -a (\la _R - \la ) - b \qquad (\hat A \ll 1).
\EE
Since the functional form is the same in the two limits
(the omitted terms $+ \dots$, however, are different), 
in the following we shall sometimes drop
the prime at the coefficients $b_i$. 

Let us first recall the scaling Ansatz of Refs.~\cite{bock,bock_proc} 
where one considers first a power-series expansion of $\hat R^2(\hat A , N_1)$
in $N_1$. In Refs.~\cite{bock,bock_proc} actually $N_2$ is used instead of 
$N_1$, but this is trivial because for any compact triangulation we have the 
fundamental relation $3N_2 = 2N_1$. Moreover, and conceptually more important,
their expansion of $\hat R^2$ is {\em not\/} done at fixed $\hat A$, but at a 
fixed discretization scale set by the average triangle area $a_0 \equiv A/N_2$ 
through the dimensionless parameter $\hat a_0 \equiv a_0/a$:
\BE
\hat R^2(\hat a_0, N_2) = N_2 c_0 (\hat a_0) +
c_1(\hat a_0) + c_2(\hat a_0)/N_2 + \dots .
\label{ps.eq:bock_N_sc}
\EE
The coefficients $c_i(\hat a_0)$ are thus defined in the
thermodynamic (infinite area) limit.
This expansion is not
based on any {\it ab initio} calculations, but has to be justified {\it a
  posteriori} by the simulation results. 
In a second step the coefficients $c_i$ are expanded in 
Refs.~\cite{bock,bock_proc} 
into a power series in $\hat a_0$ as 
\begin{eqnarray}
c_0 &=& c_0^{(0)} + \hat a_0 c_0^{(1)} + \dots ,\\
\label{ps.eqn:c_1} c_1 &=& c_1^{(0)} + \hat a_0 c_1^{(1)} + \dots ,\\
c_2 &=& \!\!(c_2^{(0)} + \hat a_0 c_2^{(1)} + \dots )/\hat a_0,
\label{ps.eqn:c_2}
\end{eqnarray}
and the continuum limit is taken by sending the discretization scale to
zero, $\hat a_0 \longrightarrow 0$.  
A comparison with (\ref{eqn:R-infty}) then yields $b_1 = c_0^{(1)}, 
b_0 = c_1^{(0)}$, and $b_{-1} = c_2^{(0)}$.
Note, that in order to make contact with the continuum result of
eq.~(\ref{ps.eq:lnZ_b}), 
$c_2$ needs to start with a divergent term $\propto 1/\hat a_0$.
Only in the combined limit 
$N_2 \longrightarrow \infty, \hat a_0
\longrightarrow 0$, this makes sense.
But because $\hat a_0$ is fixed, and not $\hat A$, 
effectively there is no control of the crossover from 
strong $R^2$-gravity scaling behavior 
($\hat A \ll 1$) to the weak $R^2$-gravity scaling behavior ($\hat A \gg 1$).
If one takes first the thermodynamic limit in (\ref{ps.eq:bock_N_sc})
then one always obtains the values of the coefficients $c_i$  
in the limit $\hat A = N_2\hat a_0 \gg 1$, hence for weak $R^2$-gravity. 
This means, $c_1^{(0)} = \GS - 2$,
and $c_2 \longrightarrow 0$, but one has to be careful to 
make the system size always large
enough to reach this limit.
If one truncates the fit at some suitable
value of $N_2$ to explore the region where $\hat A \ll 1$, then
the continuum limit can only be taken at finite $N_2$. Because the
results of Ref.~\cite{bock_proc} were
obtained for very small $N_2$, 
finite-size effects can become important. 
Even worse, because the coefficients $c_i^{(j)}$ in the 
expansion (\ref{ps.eqn:c_1}) are constants, how can $c_1^{(0)}+2$ change 
from $\GS$ to $\GS'$, as is claimed in Ref.~\cite{bock_proc}? 
These subtleties make it appear very unlikely that one can unambiguously
extract $\GS'$ in this way. 

We therefore 
suggested in Refs.~\cite{hj95a,hj95b}
a new approach which is in spirit much closer to the continuum
analysis of Ref.~\cite{kawai_r2} as it uses $\hat A$ as the distinguishing 
control parameter between weak and strong $R^2$-gravity.
Expanding $\hat R^2(\hat A,N_2)$ at constant $\hat A$ we obtain
\BE
\hat R^2(\hat A, N_2) = N_2 d_0(\hat A) +
d_1(\hat A) + d_2(\hat A)/N_2 + \dots .
\label{eq:our_N_sc}
\EE 
The next step is to expand the coefficients $d_i$ as a power series
in $\hat A$.
The coefficient $d_1$ carries all the necessary information
to extract the string susceptibilities.
A comparison with (\ref{eqn:R-infty}) and (\ref{eqn:R-inftyb}) yields for weak 
$R^2$-gravity 
\begin{equation}
d_1(\hat A) =  b_1 \hat A  + \GS - 2 + {\cal O}(1/\hat A) 
\qquad (\hat A \gg 1), 
\label{eq:gg1}
\end{equation}
and for strong $R^2$-gravity
\begin{equation}
d_1(\hat A) = S_c/\hat A  + \GS' - 2 +  b'_1 \hat A  + {\cal O}(\hat A ^2)
\qquad (\hat A \ll 1).
\label{eq:ll1}
\EE
If we plot $d_1$ versus $\hat A$ we thus expect to see a linear behavior 
for very large $\hat A$, and a divergent behavior for small $\hat A$, from 
which we can extract $\GS$ as well as $\GS'$. 
The appearance of the classical action in (\ref{eq:ll1}) is easy to 
understand. Actually, for any regular triangulation with coordination number
$q$ of arbitrary topology we find from the Euler relation 
$\delta _i = 4\pi (1-g) /q$, and
$A_i = A/q$, therefore 
$a R^2 = a \sum \delta _i^2/A_i = 16\pi^2 (1-g)^2/\hat A = S_c/\hat A$. 
For small $\hat A$ this will be the dominant term.

As far as the determination of $\GS$ is concerned ($\hat A \gg 1$), the 
difference between our 
method and that of Ref.~\cite{bock} appears as a subtle interchange of the
order in which the thermodynamic and continuum limits are taken. We first 
take the continuum limit ($N_2 \longrightarrow \infty$) for fixed $\hat A$, 
and then the thermodynamic limit ($\hat A \longrightarrow \infty$), whereas in
(\ref{ps.eq:bock_N_sc})  
first the thermodynamic limit is taken for fixed $\hat a_0=\hat A/N_2$, 
($N_2 \longrightarrow \infty$, $\hat A \longrightarrow \infty$) and 
then the continuum limit ($\hat a_0 \longrightarrow \infty$) is performed.
For $\GS'$ our procedure is basically the same as before 
(again with fixed, but now very
small $\hat A$), whereas the procedure of Ref. \cite{bock_proc} exhibits the
problems already mentioned earlier.

Because $\hat R^2$ in (\ref{eq:our_N_sc}) 
becomes infinite in the continuum limit $N_2 \longrightarrow \infty$, 
it was suggested
to add a non-scale invariant part $q_{ij}^\alpha$ to the measure and fine-tune
the exponent $\alpha$ such that it cancels the divergent term $d_0$. However, 
a trial simulation showed within error bars no change in the relevant 
coefficient $d_1(\hat A)$ \cite{hj95b}, justifying a posteriori the use of 
the simple computer measure.
%
                     \section{SIMULATION}
%
One of the objectives of the present simulations of randomly triangulated
spheres was a comparison with previous results for fixed, non-regular
triangulations \cite{bock,hj95b}. In these studies the fixed triangulation 
of the sphere was realized as the surface of a three-dimensional cube where
six exceptional vertices have coordination number four, whereas the rest of 
them have coordination number six, see Fig.~\ref{cube_fig}. The number of 
vertices is given by $N_0 = 6(L-1)^2 + 2$, where  $L$ is the edge length of 
the cube which in  our study \cite{hj95b} was varied
from $L= 7$ up to $55$.

The randomly triangulated spheres were constructed according to
the standard Voronoi-Delaunay procedure \cite{ranlat}.
For each lattice size we generated four different realizations (copies) 
to test how sensitive physical quantities depend on the randomly chosen
realizations. In order to allow an easy comparison with our previous
results for the triangulated cube, we chose the number of sites again as
$N_0 = 6(L-1)^2 + 2$, with $L$ varying from 7, 10, 15, \dots to 55 in units 
of 5. This corresponds
to $N_0 = 218$ -- 17\,498 lattice sites,
or $N_1 = 648$ -- 52\,488 link degrees of freedom, 
or $N_2 = 432$ -- 18\,252 triangles.
To update the link lengths we used a standard multi-hit
Metropolis algorithm with a hit rate ranging from 1,\dots ,3.
In addition to the usual Metropolis procedure a change
in link length is only accepted, if the links of a triangle fulfill the
triangle inequalities. 

The area $A$ was kept fixed during the update at its initial value 
$A = \sum_i A_i = N_2/2$ in order to simulate the delta-function constraint 
in eq.~(\ref{eq:1}). In principle we would need to rescale all links during 
each link update, amounting in a non-local procedure. However, due to the 
scaling properties of the partition function, this can be absorbed in a simple
scale factor in front of the $R^2$-term. To avoid round-off errors we 
explicitly performed a rescaling after every full lattice sweep. Notice, that 
our simulation procedure is technically different from the methods employed in 
Refs.~\cite{bock,bock_proc}.

As in our previous study of a fixed cubic triangulation \cite{hj95b} we ran 
two sets of simulations. In the first set we employed the method of simulating
at constant $\hat a_0$, using the 8 values of $1/\hat a_0 = 2a$ = 10, 20, 40, 
80, 160, 320, 640, and 1280. For the last four values we performed in 
addition to the 11 simulations with $N_0 = 218$ -- 17\,498 also runs at very 
small lattice sizes of $N_0 = 26$, 56, 98, and 152, or equivalently $L = 3$, 
4, 5, and 6, to facilitate the comparison to Ref.~\cite{hj95b}.

The second set of simulations consists of runs at constant $\hat A$ at the
16 values of $\hat A = 9126/a$ with $a = 5$, 10, 20, 40, 60, 80, 120,  160, 
240, 320, 480, 640, 800, 960, 1120, and 1280, which cover roughly the range of
$\hat A = 7 - 1\,800$. Again, for each value of $\hat A$, all 11 lattice sizes 
in the range $N_0 = 218$ -- 17\,498 were simulated. Here we took the majority
 of our data, to refine our scaling analyses for $\GS$ and $\GS'$.

For each run on the four copies we recorded between 10\,000 and 50\,000 
measurements of the curvature square $R^2 \equiv \sum _i R^2_i$ on every 
second to fourth MC sweep. The statistical errors for each copy were computed
using standard jack-knife errors on the basis of 20 blocks. The integrated 
autocorrelation time $\tau _{R^2}$ of $R^2$ was usually in the range of 
5 -- 10. As the final statistical error in the average over the four random 
lattice realizations we used the standard root mean square deviation 
(which usually was of about the same size as the statistical error for each 
realization).
%
                     \section{RESULTS}
%
%
                      \subsection{Results at fixed $\hat a_0$}
We first begin with an analysis of the raw data for 
$\hat R^2 /N_2$ which was obtained for the four different realizations of 
the randomly triangulated sphere in simulations with fixed $\hat a_0 = 1/2a$. 
The data for selected values of $\hat a_0$ and $N_2$ can be found in 
Table~\ref{tab1}.
One can see that the difference in $\hat R^2$ between the copies varies only 
little with system size and value of $\hat a_0$ for the system sizes with 
$L=7$ and $L=15$, where we took 10\,000 measurements. Only for the 
lattice size $L=40$, where we performed 50\,000 measurements, the 
difference between copies is larger than their statistical error. For this 
lattice size one can also observe that the difference between copies decreases 
as $\hat a_0$ tends to zero.

In Table \ref{tab1} we listed for comparison also our earlier
data for $\hat R^2$ obtained on the surface of a cube. To make the digestion
of the data easier, we plotted in Fig.~\ref{fig2} the difference 
$\Delta \hat R^2 \equiv \hat R^2_{\rm VD} - \hat R^2_{\rm cube}$ 
of the average $\hat R^2$ on the random Voronoi-Delaunay (VD) spheres and the 
cubes. We see that the difference $\Delta \hat R^2$ depends in a quite 
complicated way on both, $N_2$ and $\hat a_0$. Most disturbing is the 
observation that the curves seem to converge to zero in the limit 
$N_2 \longrightarrow 0$, and not, as one naively would expect, 
as $N_2 \longrightarrow \infty$. This shows that 
even in the thermodynamic limit the curvature data depends on the kind of
triangulation, hence is not universal. 
Even if one sends the discretization scale $\hat a_0$ to zero, it is far from
being obvious that one obtains a unique continuum value for $\hat R^2$.

To finish our comparison for fixed $\hat a_0$, we also performed a scaling
analysis on the data set 
obtained on the random sphere according to
eq.~(\ref{ps.eq:bock_N_sc}) to obtain an estimate for $\GS' = 2 + c_1$.
As already described in our earlier work \cite{hj95b} one needs to go to very
large values of $1/\hat a_0$ and small values of $N_2$ in order to obtain fits
of sufficiently high quality. The smallest value of $N_2$ we used in this
context was $N_2 = 48$. In Table \ref{t_1_v_h} we show the results for $c_0,
c_1$, and $c_2$ together with the degrees of freedom of the fit ($dof$) and
the $\chi^2$ value, for our largest values of $a$. The random sphere values
utilize the same number of $dof$ as the cube values from Ref.~\cite{hj95b} 
listed above, and below we show the fit results for somewhat more
acceptable values of $\chi^2$. The results for $c_1$ show that, if one uses the
same number of data points on the same lattice sizes, then the value of $c_1$
on the random spheres
is more stable and we get an average $c_1 = -1.0(1)$, resulting in $\GS' =
1.0(1)$. Also the values for $\chi^2$ are smaller, but still not very
satisfying. If one discards even more data points on the larger lattices, one
finally gets to a more acceptable $\chi^2$, but for the cost of having only 
a few remaining degrees of freedom. The value of $c_1$ on the random spheres 
then seems to approach again the theoretically expected value of $c_1 = -2$ as 
$\hat a_0 \longrightarrow 0$ ($a \longrightarrow \infty$), but the whole 
fitting procedure is still not convincing, see also Fig.~\ref{gs'.fig}. The 
value for $c_2\hat a_0$ comes out almost invariantly close to the expected 
value of $16\pi^2 \approx 157.91$, hence cannot serve as a test for the 
quality of the fits.
We conclude again for this section that the estimates for $\GS'$ obtained in
this fashion should not serve as a test for quantum Regge calculus, as was
advocated in \cite{bock,bock_proc}.
%
%
                      \subsection{Results at fixed $\hat A$}
Let us as well begin with a comparison of the values 
for $\hat R^2 /N_2$ obtained 
for the different
realizations of the random sphere which are compiled in Table~\ref{tab2}.
Noteworthy is that for the larger system sizes the
different realizations of the random sphere assume almost the same value within
their statistical error, because this time the statistics on the largest
lattice size was lower (10\,000 measurements) compared to the statistics
on the two smaller lattice sizes (50\,000 measurements). Also one can
observe a tendency that the difference between the copies decreases as $\hat A$
decreases. For small $\hat A$ the different copies agree with their average
within their statistical error.

If one compares now the average of the results for the random spheres with
the values obtained for the cube, one observes that the difference in
$\hat R^2$ depends, as one should expect, on the lattice size, such that the
difference between the two triangulations decreases as $N_2$ increases (which
here corresponds to the continuum limit at fixed $\hat A$). 

However, the difference in $\hat R^2$ also depends on $\hat A$ such that for
small $\hat A$ the value of $\hat R^2$ is larger on the cube than on the 
randomly triangulated sphere ($\Delta \hat R^2 < 0$), whereas for large 
$\hat A$ the value of $\hat R^2$ is smaller on the cube 
($\Delta \hat R^2 > 0$). This feature can best be inspected 
in Fig.~\ref{fig2}, where we plotted again $\Delta \hat R^2$ versus $1/N_2$. 
Strictly speaking this shows that $Z$ of eq.~(\ref{eq:1}) and $\hat R^2$ of
eq.~(\ref{eq:our_N_sc}) depend also on the way the manifold is triangulated,
hence $\hat R^2$ is a non-universal quantity. This time, however, the data
clearly shows that, if one sends $N_2$ to infinity, one really goes to the 
continuum limit in which both triangulations should give the same result for 
$\hat R^2$.

If this feature remains true for the coefficients $d_1$ is far from 
being obvious. 
Especially for large 
$\hat A$, i.e., the region which determines $\GS$, the values for $d_1$
depend on the triangulation, and one can expect universal behavior 
only for very large lattices. This effect is clearly not visible among 
the four copies of the random triangulations, therefore we
consider them to be the preferred choice of discretization scheme. 

As described in Sec.~3, the estimation of the exponents $\gamma_{\rm str}$
and $\gamma'_{\rm str}$ from our raw data for $\hat R^2(\hat A, N_2)$ 
requires a two-step fitting procedure. In the first step one
extracts the values of $d_1(\hat A)$ from fits to the FSS behavior 
according to eq.
(\ref{eq:our_N_sc}). 

In our previus short communication we used linear two-parameter fits, based
only on $d_0$ and $d_1$. Since the curves
turned out to be considerably curved we tried this time 
to account for these corrections to 
the leading FSS behavior by including also the $d_2$ correction of
eq. (\ref{eq:our_N_sc}), which improved the quality of the fits considerably.
The values for the $\chi^2/dof$, however, were still considerably larger than
one. A 
closer look at the error bars in $\hat R^2$ revealed however, that they are
probably underestimated, since we had only four random realizations available,
and 
sometimes the thermal average of a single copy was already comparable to the
standard deviations over the four copies of the random sphere.
This makes it very difficult to decide self-consistently which fit Ansatz is
the proper choice.
For all values of 
$\hat A$ the quality of the fits can be inspected in Fig.~\ref{fig:d1fit}.
All data points for $d_0$, $d_1$, and $d_2$, together with the value for
$\chi^2/dof$ and the number of
degrees of freedom of the fits are collected in 
Table~\ref{tab:4}.
The values of $d_1$ as a function of all available $\hat A$ is 
shown in Fig.~\ref{fig3}.

In the next step the small-$\hat A$ and large-$\hat A$ regimes have to
be treated separately. For small values of $\hat A < 120$ we fitted 
$d_1(\hat A)$ according to the Ansatz (\ref{eq:ll1}), which yields 
\begin{eqnarray}
S_c &=& 151(5), \label{eq:S_c_MC}\\
\GS' &=& -0.8(5) \label{eq:gs'_MC},
\end{eqnarray}
with a $\chi^2/dof = 3$, see Fig.~\ref{fig4}.
This result is still compatible with the theoretical prediction 
\cite{kawai_r2} $S_c = 16\pi ^2 \approx  158$ and $\GS' = 0$. Compared with 
our previous results on the triangulated cube ($S_c = 187(104)$,
$\GS' = 5.1(6.6)$) \cite{hj95b}, the precision of 
(\ref{eq:S_c_MC}) and (\ref{eq:gs'_MC}) is improved.
We attribute this not only to the somewhat better scaling behavior on random
triangulations and the much higher statistics (we average over 4 realizations,
and the runs for each realization have in most cases a 5 times higher 
statistics than those on the cube), but also and more important 
to the fact that in the 
present study we have much more data point available in the interesting 
small-$\hat A$ regime.

However even with the enlarged data set it turned out to be very difficult to
control systematic errors caused by the uncertainty of the proper fit
Ansatz. For example, we also tested a 2-parameter fit where the classical
action was hold fixed at its theoretically predicted value. For this fit we
obtained a value of $\GS' = -1.5(2)$ with almost the same 
$\chi^2 /dof \approx 3$. 
This shows that the systematical errors are larger then the statistical 
errors, and this is exactly the problem which makes the interpretation of
the results so difficult.

For large $\hat A$ we can employ Ansatz (\ref{eq:gg1}). However, there are 
only three data points with a sufficiently large $\hat A$ available, so no
trustworthy estimate can be obtained. A trial
linear fit through the last three points yields $\GS = -22(2)$ with a total
$\chi ^2 = 7.4$. It is not clear, 
however, if we are already in the asymptotic regime, which might set in at 
much larger $\hat A$. Another difficulty is that one is not interested in 
the slope, but in the intersection of the fit with the $y-$axis, which is far
from the location of the points used in the fit and is thus numerically very
unstable. As a third large problem we have that the $d_1$
estimates from the first fit depend very sensitively on the fit range which is
used for the first fits, meaning that one has no good control over the FSS
effects.

The systematic uncertainty 
of our value for $\GS$ is therefore hard to estimate, but it is surely an
order of magnitude larger than the quoted statistical error. However it is interesting
to note that our estimate for $\GS$ is much too negative, which is just opposite to 
what has been claimed in Ref. \cite{bock} by using their conceptually 
different FSS method.
%
                   \section{CONCLUSIONS}
%
The quantitative difference in $\hat R^2$ between the non-regular 
triangulation
and the random triangulation of the sphere depends on both, $\hat A$ and
$N_2$. In this way one will obtain on the usually used system sizes different
values of $d_1(\hat A)$. The difference seems to be negligible for small
values of $\hat A$, so that $\GS'$ can be consistently obtained on both, the
cube and the randomly triangulated sphere. However,
the difference becomes important for large values
of $\hat A$, and is thus a potential problem for the determination of $\GS$.
Because the string susceptibilities are universal quantities, one would
expect that the same holds true for the coefficients $d_1$. However, then the
continuum limit $N_2 \longrightarrow \infty$ is approached with different
speeds on the various lattice realizations.

Random triangulations appear to be a good alternative for topologies where no
large regular triangulations exist. 
They show good scaling behavior, and the differences between different copies
of the same area
decrease as the system size increases. In this way they can
provide a ``typical'' lattice for the evaluation of expectation values with 
the partition function of eq.~(\ref{eq:1}). 

Our FSS method of fitting at constant values of $\hat A$ 
gives results for $\GS'$ which are still compatible with the
theoretical prediction. In contrast to Ref. \cite{bock_proc}
we employ a consistent FSS scheme and also much larger lattices. It
would be interesting to test if contrary to \cite{bock_proc} also for
topologies of higher gender the theoretical expectations 
for $\GS'$ can be confirmed.

Due to the few data points, only a crude estimate for $\GS$ could be
obtained which, however, appeared to be too negative compared to the KPZ
theory. 
This is exactly opposite to what has been found in \cite{bock} with a
different FSS Ansatz. We
attribute this discrepancy to their method which, in our opinion \cite{hj95b},
bears conceptual problems for large values of $\hat A$. 
It is, however, unclear, if our system sizes are already in the
asymptotic scaling regime, so that the potential danger of systematic errors 
is still very large. 

To summarize, we feel that on the basis of all estimates
which have been obtained for $\GS$ and $\GS'$ so far there is no contradiction
to the theoretical predicted values. 
This is, however, mainly due to the large uncertainties
in the estimated string susceptibilities. Our studies clearly show
that on the basis of measurements for the string susceptibilities it is
neither fair to claim a success nor a failure of quantum Regge 
calculus, as has been done before.

Although our FSS method for $\hat A$ should conceptually work, it seems very
unlikely that the accuracy of the estimates for $\GS$ (and the same holds to
some extent for $\GS '$) can be improved in a 
reasonably sized computer simulation
study to serve as a stringent test for the Regge method. 
The lattice sizes which are needed to reduce corrections to scaling
appear to be huge. The problem is that the susceptibility exponents
are only subdominant corrections to the leading large area behavior, and as
such are very difficult to estimate through FSS studies. It would be definitely
more efficient to have a more direct way of measuring $\GS$ (or $\GS '$), as
it is done for example in the dynamical triangulation method \cite{q-regge}.
\section*{ACKNOWLEDGMENTS}
W.J. thanks the Deutsche Forschungsgemeinschaft (DFG) for support
through a Heisenberg Fellowship. He also acknowledges partial support by the
German-Israel-Foundation (GIF) under contract No. I-0438-145.07/95.
The numerical simulations were performed on a T3D parallel computer of
Konrad-Zuse-Zentrum f\"ur Informationstechnik Berlin (ZIB) and on other
computers of the North German Vector Cluster (NVV) in Berlin and Kiel 
under grant bvpf01.
\vspace{-0.2cm}
            
\newpage
%
%
\begin{table}[tb]
 \begin{center}
  \begin{tabular}{|c|r|r|r|}
\hline
\multicolumn{1}{|c|}{}            &
\multicolumn{1}{|c|}{$\hat R^2(\hat a_0=1/10,N_2)/N_2$}  &
\multicolumn{1}{c|}{$\hat R^2(\hat a_0=1/160,N_2)/N_2$}  &
\multicolumn{1}{c|}{$\hat R^2(\hat a_0=1/1280,N_2)/N_2$} \\
\hline
\multicolumn{1}{|c|}{}            &
\multicolumn{3}{|c|}{$N_2 = 432$, $L= 7$}         \\
\hline
 copy 1    &  0.2483(5) &  0.3818(6) &  1.3306(6) \\
 copy 2    &  0.2472(4) &  0.3826(7) &  1.3315(6) \\
 copy 3    &  0.2479(5) &  0.3831(4) &  1.3307(4) \\
 copy 4    &  0.2474(5) &  0.3822(6) &  1.3318(6) \\
 average   &  0.2477(3) &  0.3824(3) &  1.3311(3) \\
 cube      &  0.2457(3) &  0.3832(6) &  1.3315(5) \\
\hline
\multicolumn{1}{|c|}{}            &
\multicolumn{3}{|c|}{$N_2 = 2352$, $L= 15$}         \\
\hline
 copy 1    &  0.2409(2) &  0.2516(2) &  0.2861(4) \\
 copy 2    &  0.2403(2) &  0.2521(3) &  0.2867(5) \\
 copy 3    &  0.2401(2) &  0.2516(3) &  0.2875(5) \\
 copy 4    &  0.2401(3) &  0.2517(3) &  0.2851(4) \\
 average   &  0.2404(2) &  0.2518(2) &  0.2863(6) \\
 cube      &  0.2352(1) &  0.2547(3) &  0.2904(5) \\
\hline
\multicolumn{1}{|c|}{}            &
\multicolumn{3}{|c|}{$N_2 = 18252$, $L= 40$}         \\
\hline
 copy 1    &  0.24063(4) &  0.24795(8) &  0.24975(5) \\
 copy 2    &  0.24063(6) &  0.24798(6) &  0.24964(4) \\
 copy 3    &  0.24001(4) &  0.24768(7) &  0.24970(6) \\
 copy 4    &  0.24039(3) &  0.24776(5) &  0.24960(7) \\
 average   &  0.24042(15)&  0.24784(7) &  0.24967(3) \\
 cube      &  0.23339(3) &  0.24789(11)&  0.25263(8) \\
\hline
   \end{tabular}
  \end{center}
 \caption[a]{
Raw data of 
$\hat R^2 (\hat a_0,N_2) = a \langle \sum _i \delta _i^2/A_i \rangle$ 
for spherical topology at fixed $\hat a_0 = 1/2a$.
\label{tab1}
}
\end{table}
%
\begin{table}[hb]
 \begin{center}
  \begin{tabular}{|r|r|r|r|r|r|}
\hline
\multicolumn{1}{|c|}{$a=1/2\hat a_0$}      &
\multicolumn{1}{c|}{$c_0$}     &
\multicolumn{1}{c|}{$c_1$}     &
\multicolumn{1}{c|}{$c_2 \hat a_0$}& 
\multicolumn{1}{c|}{$dof$}    &
\multicolumn{1}{c|}{$\chi ^2$}     \\
\hline
\multicolumn{6}{|c|}{Cube} \\
\hline
  80 &  0.2530(4) &  -2.0(2) &   158.42(6) & 3 & 11.4 \\
 160 &  0.2520(2) &  -1.7(6) &   158.14(2) & 5 & 16.4 \\
 320 &  0.2522(2) &  -1.6(1) &   158.04(2) & 6 & 17.0 \\
 640 &  0.2535(2) &  -2.1(1) &   158.00(1) & 6 & 75.3 \\
\hline
\multicolumn{6}{|c|}{random Sphere} \\
\hline
  80 &  0.2489(2) &  -1.0(1) &   158.09(4) & 3 &  7.4 \\
 160 &  0.2486(1) &  -0.9(1) &   158.02(2) & 5 & 18.5 \\
 320 &  0.2490(1) &  -0.9(1) &   157.99(2) & 6 &  7.9 \\
 640 &  0.2493(1) &  -1.1(1) &   157.96(1) & 6 & 21.7 \\
\hline
\multicolumn{6}{|c|}{random Sphere with acceptable $\chi^2$} \\
\hline
  80 &  0.2497(5)  &  -1.2(2) &   158.15(5) & 2 &  3.6  \\
 160 &  0.2498(5)  &  -1.2(2) &   158.07(3) & 2 &  4.7  \\
 320 &  0.2501(7)  &  -1.3(3) &   158.02(3) & 2 &  1.4  \\
 640 &  0.2514(7)  &  -1.8(3) &   157.99(2) & 2 &  1.1  \\
\hline
   \end{tabular}
  \end{center}
 \caption[a]{
Fit results for the Ansatz 
$\hat R^2(\hat a_0,N_2)/N_2 = c_0 + c_1/N_2 + c_2/N_2^2$ at fixed $\hat a_0$. 
We discarded successively the largest lattices until we obtained a reasonable
total $\chi^2$ of the fit. The number of degrees of freedom for the fit is 
denoted by $dof$. For small $\hat A$ (i.e. large $a$) we expect 
$c_2 \hat a_0 = S_c = 16\pi^2 \approx 157.91$.
\label{t_1_v_h}
} 
\end{table}
%
\begin{table}[tb]
 \begin{center}
  \begin{tabular}{|c|r|r|r|}
\hline
\multicolumn{1}{|c|}{}            &
\multicolumn{1}{|c|}{$\hat R^2(\hat A = 1825, N_2)/N_2$} &
\multicolumn{1}{c|}{$\hat R^2(\hat A = 114.1, N_2)/N_2$} &
\multicolumn{1}{c|}{$\hat R^2(\hat A = 14.25, N_2)/N_2$} \\
\hline
\multicolumn{1}{|c|}{}            &
\multicolumn{3}{|c|}{$N_2 = 972$, $L= 10$}         \\
\hline
 copy 1    &  0.1913(2) &  0.2410(2) &  0.2578(2) \\
 copy 2    &  0.1836(2) &  0.2401(2) &  0.2576(2) \\
 copy 3    &  0.1874(2) &  0.2415(2) &  0.2581(2) \\
 copy 4    &  0.1897(2) &  0.2413(2) &  0.2580(2) \\
 average   &  0.1880(17)&  0.2410(4) &  0.2579(2) \\
 cube      &  0.1585(2) &  0.2364(2) &  0.2594(5) \\
\hline
\multicolumn{1}{|c|}{}            &
\multicolumn{3}{|c|}{$N_2 = 10092$, $L= 30$}         \\
\hline
 copy 1    &  0.2359(1) &  0.2469(1) &  0.2495(1) \\
 copy 2    &  0.2359(1) &  0.2472(2) &  0.2499(2) \\
 copy 3    &  0.2368(1) &  0.2471(1) &  0.2497(1) \\
 copy 4    &  0.2361(1) &  0.2469(1) &  0.2496(1) \\
 average   &  0.2362(2) &  0.2470(1) &  0.2497(2) \\
 cube      &  0.2261(1) &  0.2470(2) &  0.2536(3) \\
\hline
\multicolumn{1}{|c|}{}            &
\multicolumn{3}{|c|}{$N_2 = 34992$, $L= 55$}         \\
\hline
 copy 1    &  0.24325(4) &  0.24828(6) &  0.24950(6) \\
 copy 2    &  0.24314(5) &  0.24829(6) &  0.24960(5) \\
 copy 3    &  0.24325(4) &  0.24832(5) &  0.24960(4) \\
 copy 4    &  0.24315(4) &  0.24831(5) &  0.24950(5) \\
 average   &  0.24320(3) &  0.24830(1) &  0.24955(3) \\
 cube      &  0.23874(5) &  0.24848(6) &  0.25202(10) \\
\hline
   \end{tabular}
  \end{center}
 \caption[a]{
Raw data of 
$\hat R^2 (\hat A,N_2) = a \langle \sum _i \delta _i^2 / A_i \rangle$ 
for spherical topology at fixed $\hat A$.
\label{tab2}
}
\end{table}
%
%
\begin{table}[tb]
 \begin{center}
  \begin{tabular}{|r|r|r|r|r|r|}
  \hline
  \multicolumn{1}{|c|}{$\hat A$} &
  \multicolumn{1}{c|}{$d_0$}     &
  \multicolumn{1}{c|}{$d_1$}     &
  \multicolumn{1}{c|}{$d_2$}     &
  \multicolumn{1}{c|}{$\chi ^2/dof$} &
  \multicolumn{1}{c|}{$dof$}   \\
\hline

 1825.200 & 0.24609(7)& -108(2) &   89803(6618) & 12.4 & 6 \\
  912.600 & 0.24715(5)&  -69(2) &   61137(4819) & 14.6 & 6 \\
  456.300 & 0.24773(4)&  -43(1) &   36802(2976) &  9.8 & 6 \\
  228.150 & 0.24834(6)&  -29(2) &   24862(3119) &  2.2 & 6 \\
  152.100 & 0.24871(5)&  -25(1) &   24293(2650) &  2.1 & 6 \\
  114.075 & 0.24882(3)&  -19(1) &   18817(2387) &  3.4 & 6 \\
   76.050 & 0.24891(3)&  -12.2(4)&   7396(493)  &  8.1 & 7 \\
   57.038 & 0.24892(3)&   -9.2(4)&   6321(454)  &  6.6 & 7 \\
   38.025 & 0.24908(3)&   -4.7(3)&   4255(305)  &  5.1 & 7 \\
   28.519 & 0.24915(2)&   -1.6(3)&    3468(424) &  6.9 & 7 \\
   19.012 & 0.24932(3)&    1.8(5)&    3491(4712)&  2.4 & 7 \\
   14.259 & 0.24932(3)&    4.9(4)&    3337(384) &  9.6 & 7 \\
   11.408 & 0.24941(2)&    8.9(3)&    2322(376) &  1.9 & 7 \\
    9.506 & 0.24955(2)&   11.4(3)&    3016(354) &  4.9 & 7 \\
    8.148 & 0.24942(3)&   15.7(4)&    1450(518) &  8.3 & 7 \\
    7.130 & 0.24963(2)&   16.5(4)&    3321(351) &  3.4 & 7 \\
\hline
 \end{tabular}
 \end{center}
 \caption[a]{
Fit results for the Ansatz $\hat R^2(\hat A,N_2)/N_2 = d_0 + d_1/N_2 + d_2/N_2^2$
at fixed $\hat A$. For the six largest $\hat A$ values we discarded the smallest 
lattice size, otherwise we used the fit over all available data points.
The number of degrees of freedom for the fit is denoted by $dof$.
\label{tab:4}
}
\end{table}
%
\clearpage
\newpage
\begin{figure}[th]
\begin{center}
 \epsfig{file=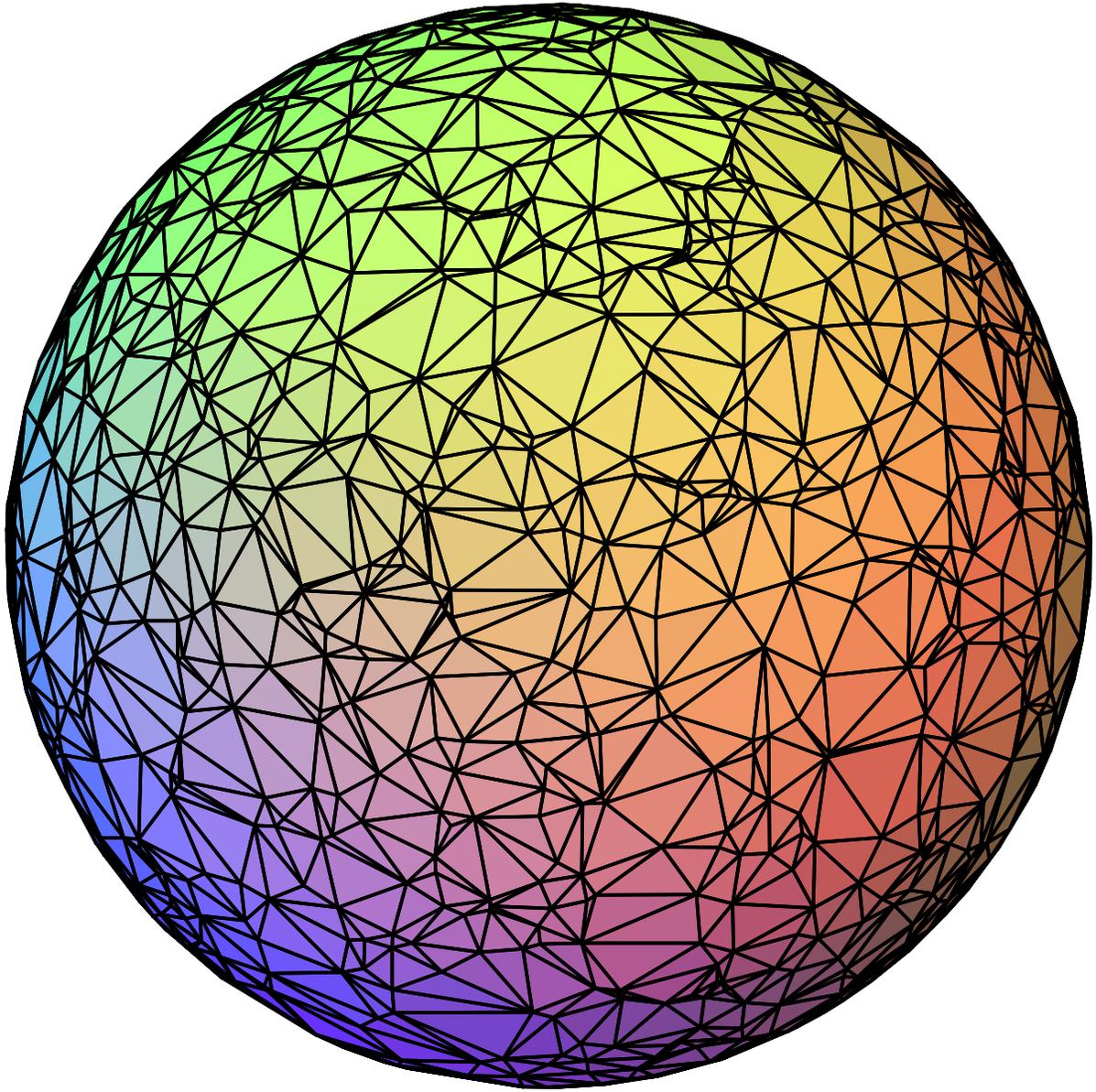,width=\textwidth}
\end{center}
\vspace{1cm}
\caption[a]{
Randomly triangulated sphere with $N_0=1500$.
\label{random_sphere.fig}
}
\end{figure}

\begin{figure}[th]
\begin{center}
\include{qdis}
\end{center}
\caption[a]{
The measured probability distribution $P(q)$ of the
 coordination numbers $q$ for the four realizations of
the spherical random lattices with $N_0 = 17\,498$
vertices (corresponding to $L = 55$).
\label{fig:q_dist}
}
\end{figure}

\begin{figure}[th]
\begin{center}
 \epsfig{file=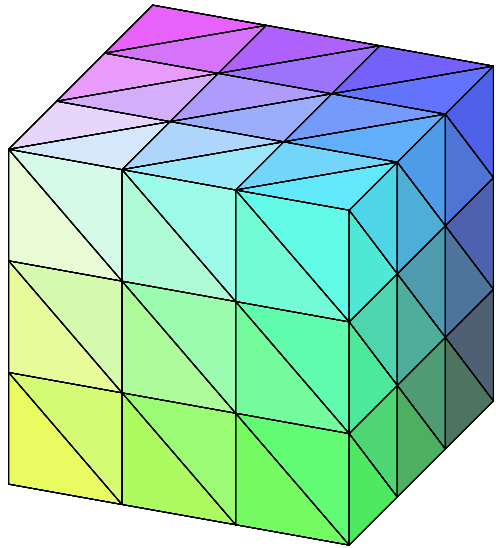,angle=-90,width=\textwidth}
\end{center}
\vspace{1cm}
\caption[a]{
The lattice realization of a spherical topology as the surface
of a three dimensional cube with $L = 4$, $N_0 = 56, N_1 = 162$, 
and $N_2 = 108$.
\label{cube_fig}
}
\end{figure}
%
%
\begin{figure}[h]
\include{fig4}
\caption[a]{
Difference $\Delta \hat R^2 /N_2$ obtained for various lattice sizes $N_2$ on the 
non-regular cube and the randomly triangulated sphere for the simulations 
with $2a = 1/\hat a_0 = const$. The curves serve only as guides to the eye.\\
\label{fig}
}
\end{figure}
\begin{figure}[h]
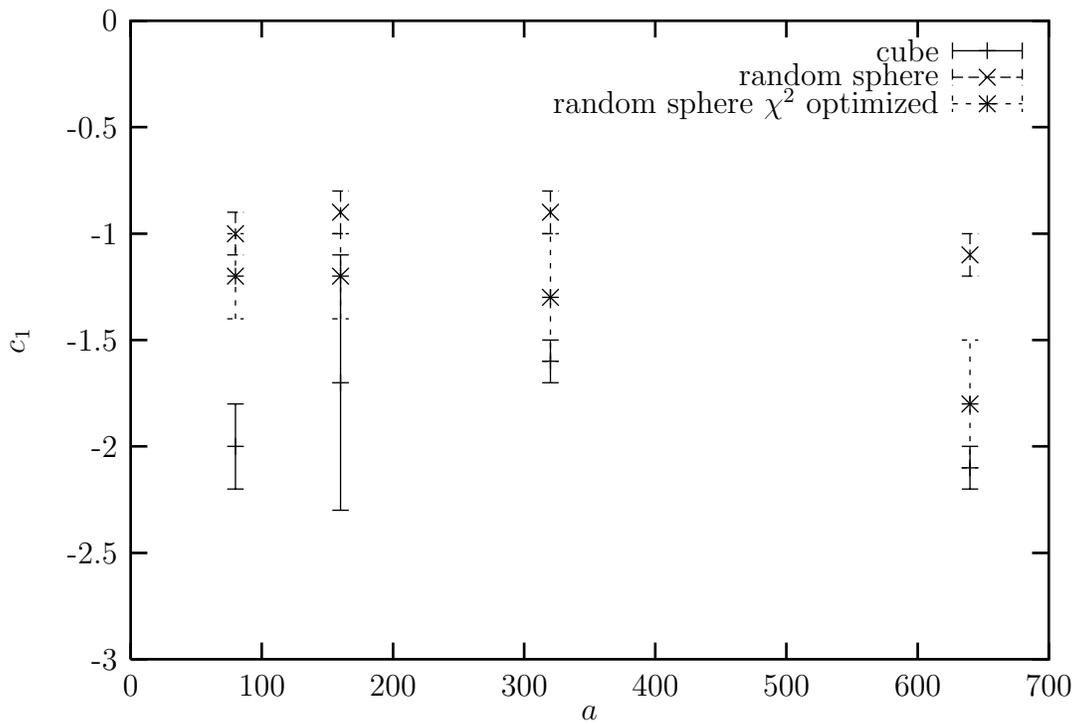

\include{gsprime}
\caption[a]{
$c_1$ plotted vs. $a$, where $c_1$ was obtained with 
eq.~(\ref{ps.eq:bock_N_sc}) on the basis of Table~\ref{t_1_v_h}.
The theory predicts a value of $c_1 = \GS' - 2  = -2$ for the
sphere. Noteworthy are the large differences between the random sphere and 
the cube data, as well as the differences due to different fit ranges.
\label{gs'.fig}
} 
\end{figure}

%
\begin{figure}[h]
\include{fig6}
\caption[a]{
Difference $\Delta \hat R^2 /N_2$ obtained for various lattice sizes $N_2$ 
on the non-regular cube and the randomly triangulated sphere for the simulations 
with $\hat A= const$.
\label{fig2}
}
\end{figure}
%

\begin{figure}[h]
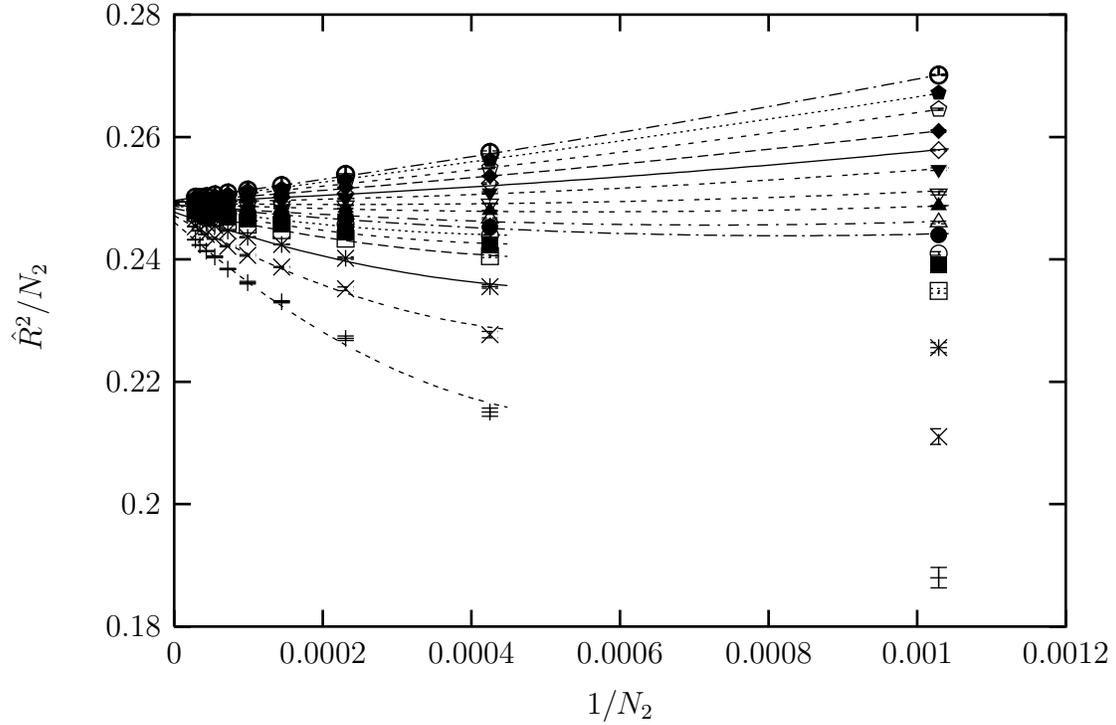

\include{fig7}
\caption[a]{
Fits of $\hat R^2 (\hat A,N_2)/N_2$ to an Ansatz 
$d_0(\hat A) + d_1(\hat A)/N_2 + d_2(\hat A)/N_2^2$, 
yielding the estimates of 
$d_1(\hat A)$ of Table~\ref{tab:4}. These values were then  
used in the second fits to extract $\GS$ and $\GS'$ 
displayed in Figs.~\ref{fig3} and \ref{fig4}. The lowest lying data points
correspond to the simulations with $\hat A = 1825.2$. The next data curves 
correspond, from bottom to top to the $\hat A$ values 912.6, 456.3, 228.2, 152.1, 
114.1, 76.1, 57.0, 38.0, 28.5, 19.0, 14.3, 11.4, 9.5, 8.1, and 7.1.
\label{fig:d1fit}
}
\end{figure}

\begin{figure}[h]
\input{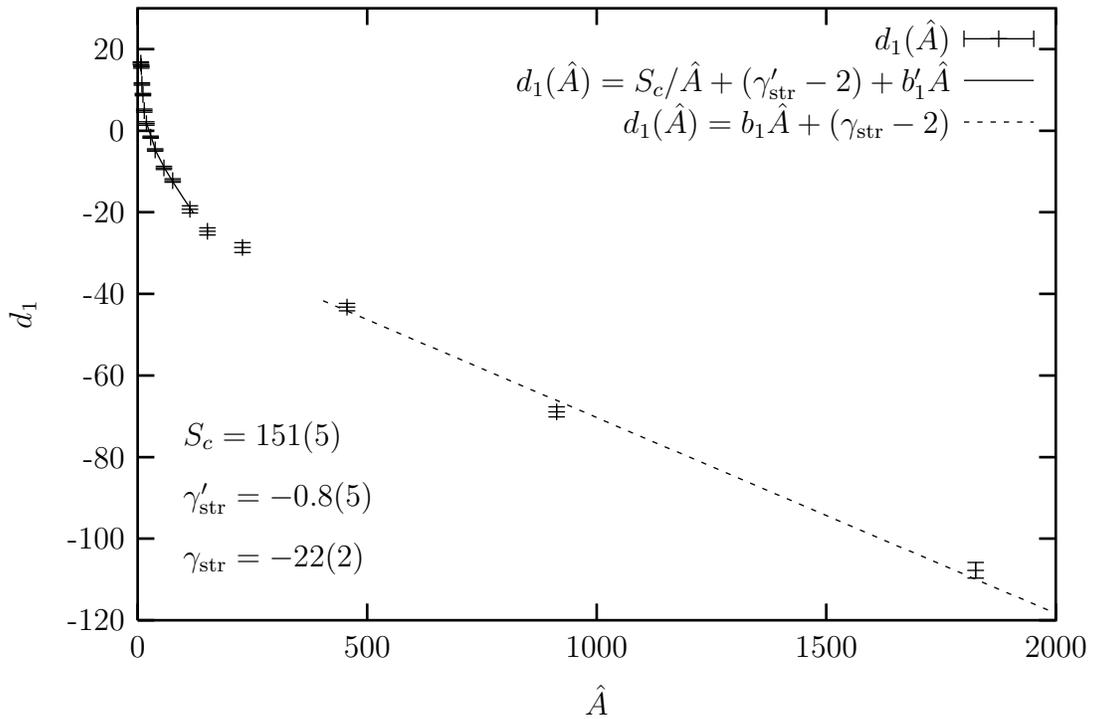}
\caption[a]{
Results for $d_1(\hat A)$ vs.\ $\hat A$ together with fits in the regimes 
$\hat A < 120$ and $\hat  A > 400$ yielding the estimates 
$\GS' = -0.8(5)$ and $\GS = -22(2)$, together with the value $S_c = 151(5)$. 
The theoretical predictions for spherical topologies
are $S_c = 16\pi^2 \approx 157.91$, $\GS' = 0$, and $\GS = -1/2$.\\
\label{fig3}
}
\end{figure}

\begin{figure}[h]
\input{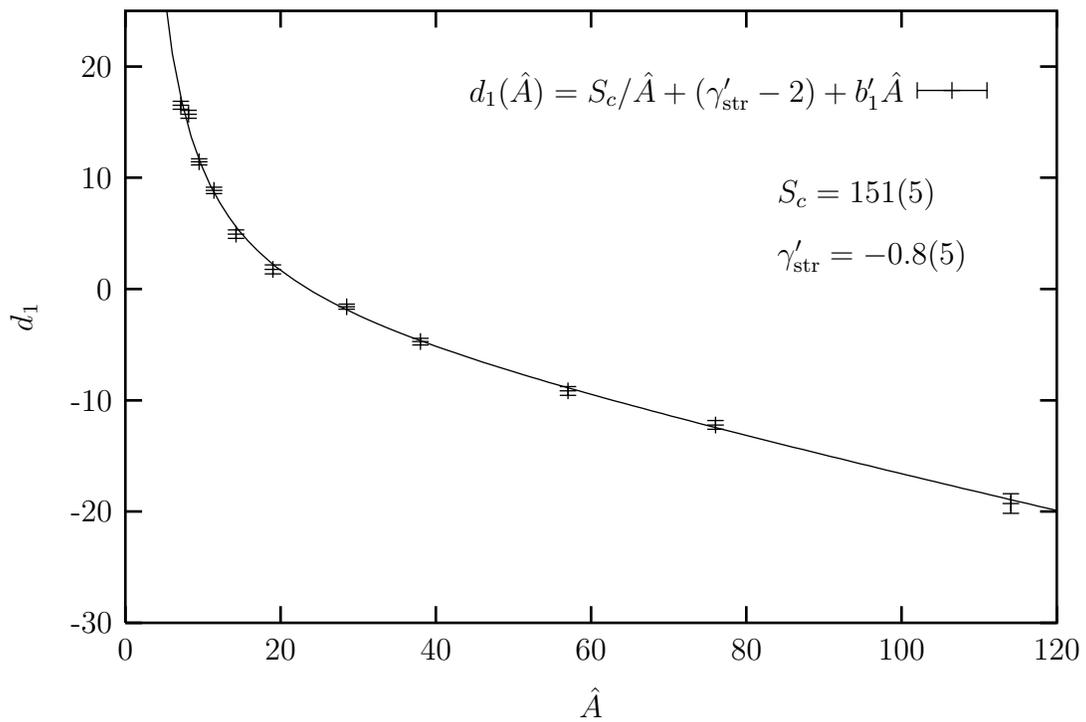}
\caption[a]{
Blow-up of Fig.~\ref{fig3} for $\hat A < 120$ yielding the estimate 
$\GS' = -0.8(5)$.\\
\label{fig4}
}
\end{figure}

\end{document}